# Multifilamentary, in-situ Route, Cu-stabilized MgB$_2$ Strands

M.D. Sumption[1], M. Bhatia[1], X. Wu[1], M. Rindfleisch[2], M. Tomsic[2], and E.W. Collings[1]

[1]LASM, Materials Science and Engineering Department,

OSU, Columbus, OH 43210, USA

[2]Hyper Tech Research, Inc. Columbus, OH 43210, USA

**Abstract**

Transport critical current densities and *n*-values were measured at 4.2 K in fields up to 15 T on 7, 19, and 37-stack multifilamentary MgB$_2$ strands made using an in-situ route. Some strands included SiC additions (particle size ? 30 nm), while in others Mg-rich compositions were used. Two basic multifilamentary variants were measured, the first had Nb filamentary barriers, the second had Fe filamentary barriers. All samples incorporated stabilizer in the form of Cu 101. Simple, one-step heat treatments were used, with temperatures ranging from 700-800?C, and times from 10-30 minutes. Transport critical current densities of 1.75 x 10$^5$ A/cm$^2$ were seen at 4.2 K and 5 T in 37 stack strands.

**Introduction**

Many groups now fabricate $MgB_2$ wires[1-13], powder-in-tube-processed (PIT) strands being very promising for current carrying applications. Typical present day strands employ Fe or Cu, with perhaps Cu-Ni or monel as the outer sheath material. There are two main variants of PIT $MgB_2$ fabrication, ex-situ[1-4], and in-situ[6,10-13]. Each of these choices has advantages and disadvantages. We focus here on in-situ powders, in particular their incorporation into multifilamentary (MF) strands.

Numerous efforts in $MgB_2$ development are ongoing, and significant progress is being made in improving basic properties such as transport $J_c$, upper critical fields, and irreversibility fields. Development of high quality MF strand geometries is now becoming important for a number of reasons; they include strain tolerance, flux jump stability, transport current stability, and AC loss reduction. The first practical superconductors, MF NbTi/Cu, were filamentarized mostly to limit flux jumping. The same need exists for $MgB_2$ because of its lower temperature applications range. The utility of filamentary subdivision as an aid to strain tolerance first became crucial in Bi-based conductors, this should be useful for $MgB_2$ composites as well. Transport current stability (and the related concept of protection) also has a long history, and requires the presence of significant amounts of highly conductive "stabilizer", usually Cu or Ag.

This last requirement has complicated the development of MF $MgB_2$ strands chiefly because the Fe barriers typically used as reaction-stopping interfaces do not have the same flow stress as the high purity Cu or Ag needed for electrical stabilization. This leads, for the most obvious strand designs, to drawing instabilities, and wire failure. One difficulty has to do with the lack of high purity Fe strip of tube and the other with the work hardening characteristics of Fe (both influencing the draw stress of the Fe). It is possible to leave out the Fe barrier, exposing the mixed powders to the Cu tube, but this allows significant Mg-Cu reaction, which "poisons" the $MgB_2$ and degrades $J_c$. Other barrier alternatives have not seemed viable, e.g., Nb, which had been thought to also cause $J_c$ degradation. Below we will demonstrate the development of simple 7, 19, and 37 stack strands using Fe and Nb protective layers, outer Cu-Ni or monel shells, and included stabilization layers (Cu in this case). The $J_c$ properties for a conductor with a high quality MF structure will be shown to be comparable to the monofilament results.

**Background**

Much of the work on MF $MgB_2$ strands has favored the use of pre-reacted powders. Dou et al.[13] reported making 4 and 16 stack strands using Fe and stainless steel as the sheath materials. The wires were square, as were many of the initial $MgB_2$ MF strands. Dou's group also made MS strands from in-situ powders, these were 7 stack (presumably round) wires[14]. The ex-situ powder route was also used by Pachla and Kovác et al. In this case both Fe and Fe/Cu sheath variants[15,16] enclosed 4, 9, and 16 filament square geometry arrays (resulting from two-axis rolling reduction techniques). The authors encountered flow stress matching problems, and the associated breakage of the Fe barrier and subsequent Cu-related poisoning. Kumakura et al demonstrated 7-stack wires in CuNi sheath material, reaching 2000 A/cm$^2$ at 4 T and 4.2 K[17]. Flükiger, et al, with attention to powder sizes, made 7-filament arrays and achieved 30,000 A/cm$^2$ at 4 T and 4.2 K [18]. Glowaki and Majoros, in another important advance, demonstrated the winding of a steel-reinforced $MgB_2$ cable[19].

Based on these early efforts, we recognized that drawing instability, stabilizer incorporation, superconductor fraction, and the usability of wire drawing (rather than rolling techniques) were areas which needed to be addressed. Below we will show some simple 7, 19, and 37 stack strands which do this. In these wires, thin layers of Fe or Nb are used to protect a further layer of Cu from the in-situ Mg + B powders. The Fe- or Nb-clad powders are then inserted into a Cu can first to facilitate drawing and later to serve as a stabilizing layer in the final wire. After some area reduction, these monofilaments are restacked as subelements into CuNi or monel cans, to provide proper flow stress matching and retain compaction during HT and cool down.

**Experimental**

*Sample preparation*

The continuous tube forming/filling (CTFF) process was used to produce the subelements for $MgB_2$/Fe composite strands [10,11]. This process, as developed at Hyper Tech Research (HTR), is depicted in Figure 1 for a restack strand. First, the powder was dispensed onto a strip of metal as it is being continuously formed into a tube. For $MgB_2$ strands the strip is commercially pure Fe (23 mm wide, 0.5 mm thick) or Nb (23 mm wide, 2 x 0.25 mm thick) strip running at about 0.4 m/min. After exiting the mill at a diameter of 5.9 mm the filled overlap-closed tube was inserted into a full hard 101 Cu tube. After drawing to the proper size, these monofilaments were then restacked round into 7, 19, or 37 subelement arrays inside of either Cu-30 Ni or monel outer tubes and then drawn to final size. A 19 stack strand is shown in Figure 2 (a) and a 37 stack strand in Figure 2 (b). For further details of this process see [10,11].

The starting 99.9% Mg powders were -325 mesh (but had an approximate top size of 20 ?m), and the 99.9% B powders were amorphous, at a typical size of 1–2 ?m. The powders were V-mixed and then run in a planetary mill. Three powder types were made: stoichiometric binary powder, stoichiometric binary powder with SiC additions, and Mg rich powders without SiC additions. SiC, when present was added during the V-mixing stage, in the ratio of 10 mole % SiC to 90 mole % of binary $MgB_2$; that is, $[(MgB_2)_{0.9}(SiC)_{0.1}]$. The SiC particles had an average diameter of about 30 nm. For one powder batch Mg was added to the initial Mg and B mixture form the ratio $Mg_{1.1}B_2$. These powders contained no SiC. Table 1 describes the MF strands and their general characteristics.

Heat treatments were then performed under flowing Ar. Ramp-up times were typically 45 min, and the samples were furnace cooled over three hours, this is denoted furnace ramp in Table 2. In some cases, the samples were inserted into a furnace which was already at temperature, and after the plateau partially pulled out, cooling over 1.5 hours, this is denoted rapid ramp in Table 2. The times and temperatures at the plateau ranged from 10-30 minutes at 700-800?C. The particular HT plateau parameters are given in Table 2.

*Measurements*

Four-point transport $J_c$ measurements were made at 4.2 K in liquid He. Measurements were made in background ?elds of up to 15 T applied transversely to the strand. The samples were 3 cm in length, with a gauge length of 5 mm. The $J_c$ criterion was 1 ?V/cm. Values for *n* were obtained by taking $E = \text{Const}(I/I_c)$ as the form for current and voltage above $I_c$, and using two points, 1 µ/cm and 10 µV/cm to define $n$. We calculated $n$ using $n = 1/[\log(I_2/I_1)]$ where $I_1$ is defined via the first criterion and $I_2$ via the second.

**Results**

One of the important results of this study is the relative uniformity of the filamentary structures. We note that the maximum size of the Mg powders were mostly kept below 20 µm (although some larger particles up to 45 µm were present in the initial batch). We note that particle size was not greatly reduced by planetary milling. When we combine this information with the fact the filament core (the $MgB_2$) diameters were approximately 150 µm, 120 µm, and 80 µm for the 7, 19, and 37 stack strands, respectively, we see that the Mg particles must be deforming significantly during drawing, relaxing the practical criterion that filament size should be 10 x the particle size of the largest elements (typically used for Bi-based and other hard particle composites). This fact will be quite important for the development of practical $MgB_2$ multifilamentary strands.

The results of transport $J_c$ measurements on all strands is shown in Figure 3. Results for 7, 19, and 37 filament strands are displayed, some with Fe barriers and others with Nb barriers. The 7 filament strand results are all quite similar, with no significant difference between the Fe and Nb-based sample results. Two of the 19 stack Nb-based strands provide the best $J_c(B)$. The 37 stack sample result is similar to those of the 7 stack strands. Comparing the results of these samples to monofilamentary strands is instructive. Values of 7 x $10^4$ A/cm$^2$ at 5 T, 2 x $10^4$ A/cm$^2$ at 8 T, and 6 x $10^4$ A/cm$^2$ have been typical monofilamentary $J_c$ values for wires made at HTR incorporating 200 nm SiC additions using an Fe chemical barrier and an outer CuNi or monel sheath. These MF results are similar to the monofilamentary results, in some cases lower, but in some cases somewhat higher. This may be due either to excess Mg or finer SiC particle additions (these are 30 nm in size).

Extracted *n*-values generally ranged from 3 to 10. No quenching was seen up to 220 A. This is the limit of our short sample probe, in any case reliable measurements are not expected for short wire samples above this current level. The absence of quenching is in strong contrast to the behaviors of earlier unstabilized samples which frequently had quench-like transitions.

**Summary and Conclusions**

Transport critical current density ($J_c$) was measured at 4.2 K in fields of up to 15 T on 7, 19, and 37-stack MF $MgB_2$ strands made using an in-situ route. In some cases SiC was included (particle size ? 30 nm), while in others Mg-rich compositions were used. Two basic MF variants were measured, the first had Nb-filamentary barriers, the second had Fe-filamentary barriers. All samples incorporated stabilizer in the form of Cu 101. Filament core sizes of approximately 150 ?m, 120 ?m, and 80 ?m were achieved for the 7, 19, and 37 stack strands, respectively. Simple, one-step heat treatments were used, with temperatures ranging from 700-800?C, and times from 10-30 minutes. $J_c$ values reached in some cases 1.75 x $10^5$ A/cm$^2$ at 5 T and 4.2 K (1?V/cm), with *n* values ranging from 3-10.

**Acknowledgements**

This work was supported by a State of Ohio Technology Action Fund Grant. Partial support by the US Department of Energy, HEP, Grant No. DE-FG02-95ER40900 is also acknowledged. Support from the Air Force under Grant No. F33615-03-C-2344 is also acknowledged.

**List of Tables**

**List of Figures**

Figure 1. Schematic of CTFF-Multifilament Process.

Figure 2. (a) 19 subelement Nb/Cu/monel MF strand. (b) 37 subelement Nb/Cu/monel MF strand.

Figure 3. Transport $J_c$ vs $B$ for various multifilamentary strands.

Table 1. Strand Specifications

| Name | Trace ID | Sub. No. | Sub. Type | Outer Can | Additive[a] | s/d | SC % | Bar % | Cu % | OD, mm |
|---|---|---|---|---|---|---|---|---|---|---|
| FeCu7MN | 434 | 7 | Fe/Cu | monel | SiC[a] | 0.41 | 15 | 15 | 28 | 0.83 |
| FeCu7CN | 435 | 7 | Fe/Cu | Cu-30Ni | SiC[a] | 0.41 | 13 | 13 | 31 | 0.83 |
| NbCu7 | 501 | 6 | Nb/Cu | monel | -- | 0.45 | 13 | 12 | 33 | 0.83 |
| NbCu19+M[b] | 516 | 18 | Nb/Cu | monel | Mg[b] | 0.28 | 17 | 14 | 24 | 1.0 |
| NbCu37+M[b] | 518 | 36 | Nb/Cu | monel | Mg[b] | 0.28 | 14 | 14 | 27 | 1.0 |

[a] 10 mole % of SiC added to 90 mole % of binary $MgB_2$ [$(MgB_2)_{0.9}(SiC)_{0.1}$]. SiC used was ~ 30 nm.

[b] Mg was added to form the ratio $Mg_{1.1}B_2$

Table 2. Sample Specifications and Heat Treatments

| Name | Trace ID | HT (?C/min) | HT ramp type |
|---|---|---|---|
| FeCu7MN-700/20 | 130.8/434 | 700/20 | Rapid |
| FeCu7CN-700/10 | 130.6/435 | 700/10 | Rapid |
| FeCu7CN-700/20 | 130.9/435 | 700/20 | Rapid |
| NbCu7-700/10 | 147.6/501 | 700/10 | Furnace |
| NbCu7-700/30 | 147.12/501 | 700/30 | Furnace |
| NbCu7-800/10 | 147.18/501 | 800/10 | Furnace |
| NbCu7-800/30 | 147.24/501 | 800/30 | Furnace |
| NbCu19-700/20 | 153.20/516 | 700/20 | Furnace |
| NbCu19-700/5 | 153.15/518 | 700/5 | Furnace |
| NbCu19-700/10 | 153.18/518 | 700/10 | Furnace |
| NbCu37-700/20 | 153.21/518 | 700/20 | Furnace |

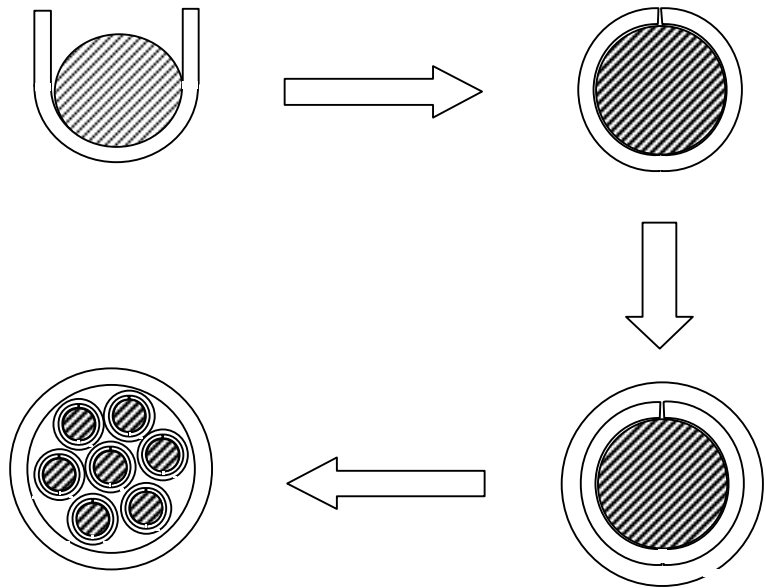

**SUMPTION Figure 1**

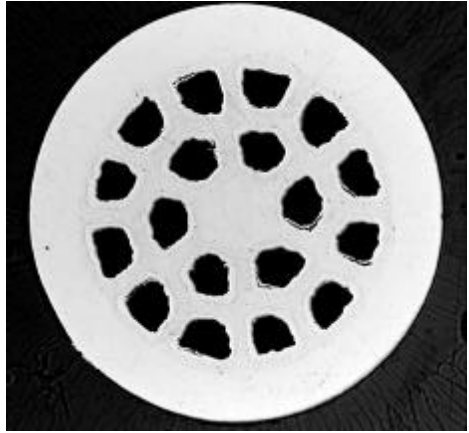

**SUMPTION Figure 2 (a)**

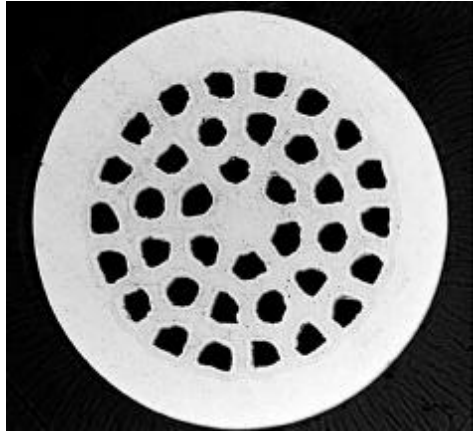

**SUMPTION Figure 2 (b)**

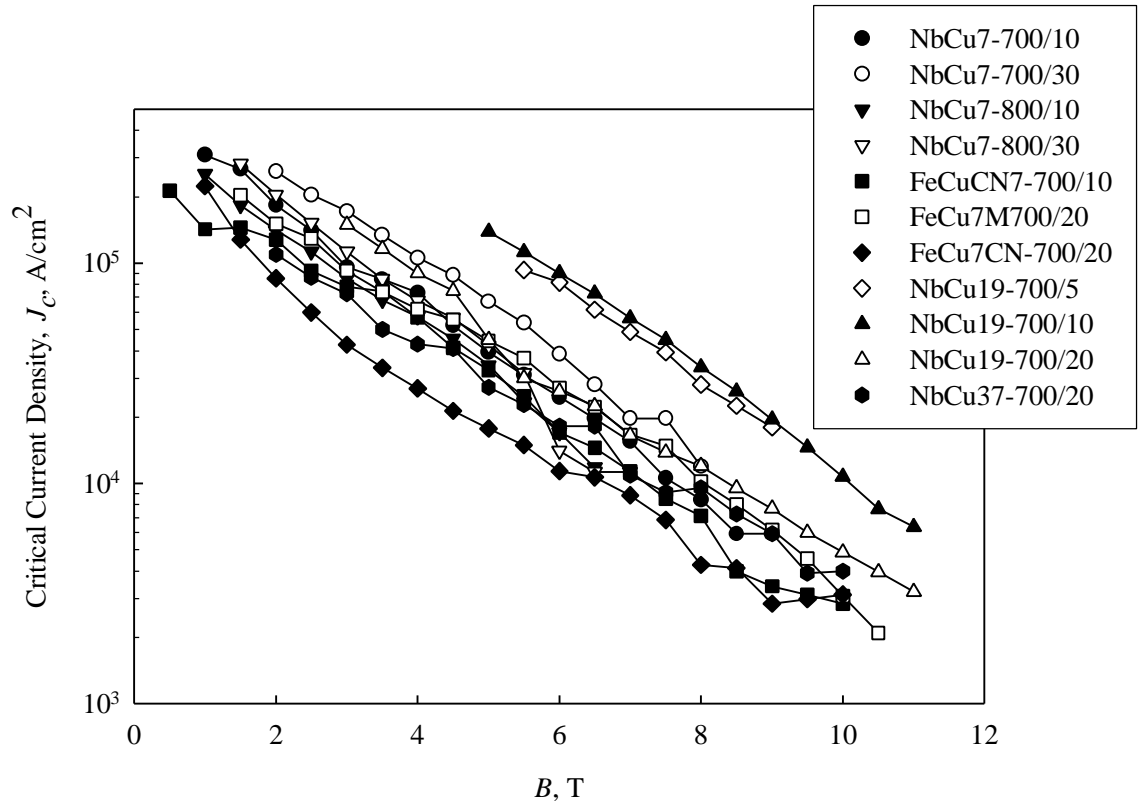

**SUMPTION Figure 3.**